\documentclass[aps,prl,twocolumn,superscriptaddress]{revtex4}
\usepackage{amssymb}
\usepackage{amsfonts}
\usepackage{graphicx}
\usepackage{CJK}
\usepackage{indentfirst}
\usepackage{amsmath}
\begin{document}


\title{Gap Structure of the Overdoped Iron-Pnictide Superconductor Ba(Fe$_{0.942}$Ni$_{0.058}$)$_{2}$As$_{2}$: A Low-Temperature Specific-Heat Study}

\author{Gang Mu}
\email[]{mugang@mail.sim.ac.cn} \affiliation{State Key Laboratory of
Functional Materials for Informatics and Shanghai Center for
Superconductivity, Shanghai Institute of Microsystem and Information
Technology, Chinese Academy of Sciences, Shanghai 200050, China}
\author{Bo Gao}
\affiliation{State Key Laboratory of Functional Materials for
Informatics and Shanghai Center for Superconductivity, Shanghai
Institute of Microsystem and Information Technology, Chinese Academy
of Sciences, Shanghai 200050, China}
\author{Xiaoming Xie}
\affiliation{State Key Laboratory of Functional Materials for
Informatics and Shanghai Center for Superconductivity, Shanghai
Institute of Microsystem and Information Technology, Chinese Academy
of Sciences, Shanghai 200050, China}
\author{Yoichi Tanabe}
\affiliation{Department of Physics, Graduate School of Science,
Tohoku University, Sendai 980-8578, Japan}
\author{Jingtao Xu}
\affiliation{World Premier International Research Center, Tohoku
University, Sendai 980-8578, Japan}
\author{Jiazhen Wu}
\affiliation{Department of Physics, Graduate School of Science,
Tohoku University, Sendai 980-8578, Japan}
\author{Katsumi Tanigaki}
\email[]{tanigaki@sspns.phys.tohoku.ac.jp} \affiliation{Department
of Physics, Graduate School of Science, Tohoku University, Sendai
980-8578, Japan}\affiliation{World Premier International Research
Center, Tohoku University, Sendai 980-8578, Japan}

\begin{abstract}
Low-temperature specific heat (SH) is measured on the post-annealed
Ba(Fe$_{1-x}$Ni$_{x}$)$_2$As$_2$ single crystal with $x$ = 0.058
under different magnetic fields. The sample locates on the overdoped
sides and the critical transition temperature $T_c$ is determined to
be 14.8 K by both the magnetization and SH measurements. A simple
and reliable analysis shows that, besides the phonon and normal
electronic contributions, a clear $T^2$ term emerges in the low
temperature SH data. Our observation is similar to that observed in
the Co-doped system in our previous work and is consistent with the
theoretical prediction for a superconductor with line nodes in the
energy gap.
\end{abstract}

 \maketitle

\section{1. Introduction}
The superconducting (SC) state of a superconductor is protected by a
energy gap. The symmetry and structure of the energy gap can be very
different in different SC materials. For the conventional
superconductors (e.g. metallic superconductors), the gap is
isotropic in the k space which is called the s-wave
symmetry~\cite{MgIrB}. In some materials (e.g. MgB$_2$), multlple
gaps have been discovered on different Fermi surfaces~\cite{MgB2}.
Highly anisotropic (the so-called d-wave) gap symmetry was confirmed
in high-$T_c$ cuprate superconductors~\cite{LSCO}. The situation is
more complicated in the iron-pnictide superconductors, because there
are typically four or five bands crossing the Fermi level.
Theoretically several candidates symmetries of the SC gap(s) were
proposed~\cite{Hirschfeld}, among which the so-called S$^{\pm}$ case
seems to be accepted widely~\cite{S-PM1,S-PM2}. On the experimental
sides, the nodeless superconductivity has been confirmed in
(K,Tl)$_x$Fe$_{2-y}$Se$_2$~\cite{KFe2Se2}. However, nodes (zero
points) have been reported in the gap(s) of LaFePO, KFe$_2$As$_2$,
and BaFe$_2$(As$_{1-x}$P$_x$)$_2$~\cite{LaFePO,KFe2As2,Ba122-P}. At
the same time, the consensus has not been reached on other systems
of iron-pnictide
superconductors~\cite{WYL,Sato,ZhGQ,MuCPL1,Chien,HDing,Hashimoto,MuPRB,HeatTransport,ARSH}.
In the electron-doped (Co- or Ni-doped) 122 system, one tendency
that the gap anisotropy becomes large and even gap nodes emerges in
the overdoped samples has been reported by different groups and
experimental
methods~\cite{penetration-depth,HeatTransport-BaCo,anneal,optical-conductivity,MuPRB2,MuJPSJ}.

Specific heat is a bulk tool to detect the quasiparticle density of
states (DOS) at the Fermi level, which can provide information about
the gap structure. The variation of the electronic SH in the SC
states ($C_{sc}$) versus temperature can be rather different for
different gap structures~\cite{review1,review2},
\begin{equation}
C_{sc}\sim \left\{\begin{aligned}
e^{-\Delta_0/k_BT},\quad\quad \text{s-wave}\\
T^2,\quad\quad\quad \text{line nodes}\\
T^3,\quad\quad \text{point nodes}\end{aligned}\right.
\end{equation}
where $\Delta_0$ is the magnitude of the energy gap. In order to
segregate the pure electron SH from the measured mixed
contributions, many methods have been
tried~\cite{MuPRB,Keimer,Ronning}. In our previous work, we reported
the clear presence of $T^2$ term in $C_{el}$ of the overdoped
Ba(Fe$_{1-x}$Co$_{x}$)$_{2}$As$_{2}$ from the raw data, giving a
more solid evidence for the line-nodal gap
structure~\cite{MuPRB2,MuJPSJ}. It is very important and necessary
to investigate more systems with other dopants to check the
universality of such a behavior.

In this paper, we studied the the low temperature SH of the Ni-doped
BaFe$_2$As$_2$ in the overdoped region. Here we also observed a
clear $T^2$ term in $C_{sc}$, being consistent with the theoretical
prediction for the line-nodal superconductors. Our result along with
the previous work indicates that it is a universal feature of the
electron doped 122 system.

\begin{figure}
\includegraphics[width=7.5cm]{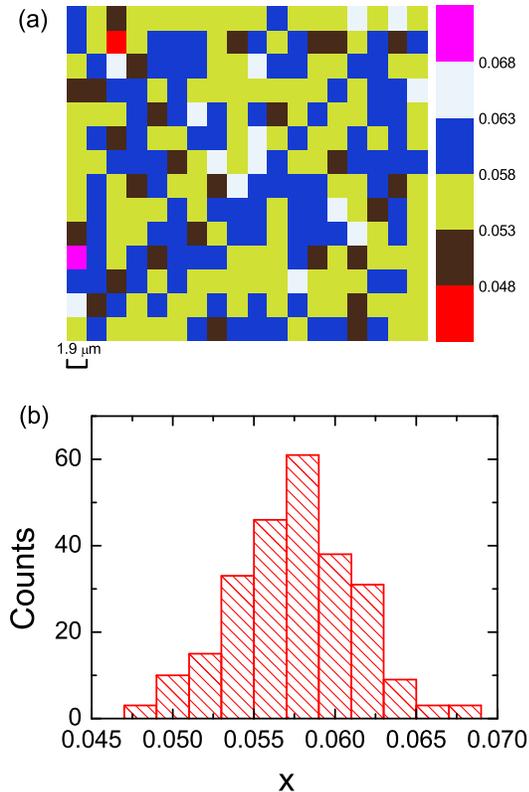}
\caption {(color online) (a) The mapping image of Ni concentration
throughout an area of 38 $\mu$m$\times$28 $\mu$m based on the EDS
quantitative results. (b) The chart of the Ni distributions
summarized from same data as (a).} \label{fig1}
\end{figure}

\begin{figure}
\includegraphics[width=8.5cm]{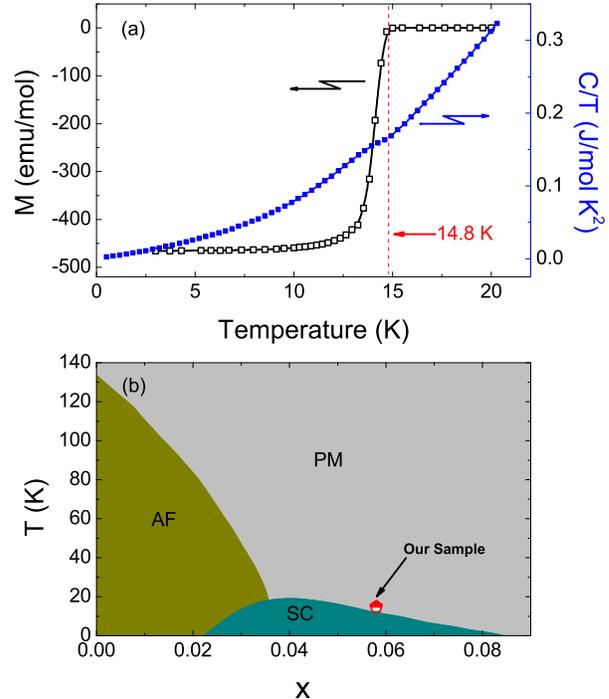}
\caption {(color online) (a) Temperature dependence of dc
magnetization (left) and specific heat coefficient $C/T$ (right) for
the sample. The magnetization data are collected with field $H$ = 10
Oe using the zero field cooling (ZFC) process. (b) Phase diagram for
the Ni doped 122 system~\cite{Ni}. The red mark shows the location
of our sample. } \label{fig2}
\end{figure}

\section{2. Materials and Methods}
The Ba(Fe$_{0.942}$Ni$_{0.058}$)$_{2}$As$_{2}$ single crystal was
grown by the self-flux method~\cite{LFang}. The as-grown sample was
annealed under high vacuum at 1073 K for 20 days, because it was
reported that the annealing process can improve the sample quality
significantly~\cite{anneal}. The sample for the SH measurement have
a mass of 2.9 mg. The actual Ni concentrations were checked and
determined by the energy dispersive X-ray spectroscopy (EDS)
measurements on a Bruker Quantax200 system. The dc magnetization
measurements were done with a superconducting quantum interference
device (Quantum Design, MPMS7). The specific heat were measured with
a Helium-3 system based on the physical property measurement system
(Quantum Design, PPMS). We employed the thermal relaxation technique
to perform the specific heat measurements. The thermometers has been
calibrated under different magnetic fields beforehand. The external
field was applied perpendicular to the $c$ axis of the single
crystal.

The distribution of the Ni-dopant on a micro-scale was investigated
by EDS measurements. The mapping image of Ni concentration and the
chart of its distributions throughout an area of 38 $\mu$m$\times$28
$\mu$m for our sample are shown in Figures 1(a) and (b). The spatial
resolution is 1.9 $\mu$m. The distribution shows a peak at about
0.058, which is very close to the average value. Consequently, this
value is taken as the actual doping level. The SC transition of the
obtained single crystal was checked by the dc magnetization and
specific heat measurements. As shown in Figure 2(a), the clear SC
transition at about 14.8 K can be seen from both the $M-T$ and
$C/T-T$ curves, indicating a high quality of the selected sample.
Figure 2(b) shows schematically the phase diagram of the present
Ni-doped system, which were reported by N. Ni et al~\cite{Ni}. It is
clear that our sample locates on the overdoped sides of the phase
diagram and no magnetic order exists in this region, which supplies
a clean platform to study the behaviors of specific heat.

\section{3. Results and Discussion}

We focus our attention on the SH data in the low temperature range
to study the low-energy excitations. Generally speaking, in a system
without magnetic order or magnetic impurities, the total SH is a
simple integration of different components,
\begin{equation}
C(T) =\beta T^3 +C_{sc}+\gamma T.
\end{equation}
The first term is the phonon SH, which is a very good approximation
for the Debye model in the low temperature region. As mentioned in
the introduction, the second term $C_{sc}$ is the electronic SH in
the SC states excited by the thermal energy. This is the most
important and concerned in our study, because its responding
behavior to temperature supplies the information of the gap
structure. The third term is quite complicated. Under zero field, it
is a residual electronic term typically coming from small amounts of
non-superconducting content in the sample or impurity scattering in
some unconventional superconductors~\cite{LSCO,MuCPL2}. With a field
higher than the lower critical field, it reflects the contribution
of the vortex states. Here we plot the raw data of SH as $C/T$ vs
$T^2$ in Figure 3. A clear feature in this figure is the negative
curvature in all the curves under different fields, rather than a
linear behavior as revealed by the dashed pink line. This behavior
is not expected for the SC gap with an s-wave symmetry or point
nodes. In the case of s-wave, the value of $C_{sc}$ is negligibly
small in the low temperature region due to the exponential relation.
For systems with point nodes, the contribution of $C_{sc}$ have the
same $T^3$-dependence as the phonon term. In both cases, the $C/T$ -
$T^2$ curves should show a linear behavior in the present low
temperature region. Intuitively, the case of gap with line nodes is
the best candidate. Consequently, we fitted our data using Eq.~(2)
based on the line nodal situation ($C_{sc} = \alpha T^2$). As
represented by the solid lines, the fitting results are shown in
Figure 3. One can see that the fitting curves coincide the
experimental data very well. This gives a direct evidence for the
presence of line nodes in the energy gap.

To further confirm the reliability of the analysis, we checked the
obtained fitting parameters carefully. We show the field dependence
of the parameters $\beta$, $\alpha$, and $\gamma$ in Figure~4 (a),
(b), and (c), respectively. The value of $\beta$ is almost
independent of field, which is reasonable because magnetic field
can't affect the phonon SH. Moreover, $\alpha(H)$ decreases
monotonously with the increase of magnetic field up to 9 T. This is
similar to that observed in overdoped
Ba(Fe$_{1-x}$Co$_x$)$_2$As$_2$, and has been attributed to the
combined effects imposed by the three-dimension dispersion of the
line nodes on the Fermi surface and the destruction of V-shape of
the density of states (DOS) at the nodes by the
field~\cite{MuPRB2,MuJPSJ}. The residual value of $\gamma$ under
zero field (denoted as $\gamma_0$) is estimated to be 0.75 mJ/mol
K$^2$, which is also comparable to the reports in overdoped
Ba(Fe$_{1-x}$Co$_x$)$_2$As$_2$ and suggests a high sample quality. A
clear increase of $\gamma$ with the field is observed. It is
difficult to describe the field dependence of $\gamma$ using a
simple formula due to the multi-band effect in the present system.
Qualitatively, $\gamma$ increases more quickly in the system with a
highly anisotropic gap. At the present stage, we could not evaluate
the information supplied by the field dependent data, since the
upper critical field $H_{c2}$ and the normal state electronic SH
coefficient $\gamma_n$ are not clear.

\begin{figure}
\includegraphics[width=9cm]{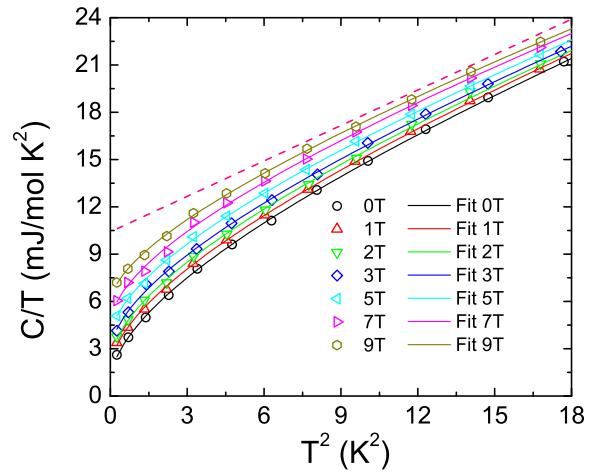}
\caption {(color online) The raw data of the SH under different
fields in the low temperature region. The data are shown in $C/T$ vs
$T^2$ plot. The solid lines display the theoretical fitting (see
text). The dashed straight line is the guide for the eyes.}
\label{fig3}
\end{figure}

\begin{figure}
\includegraphics[width=8cm]{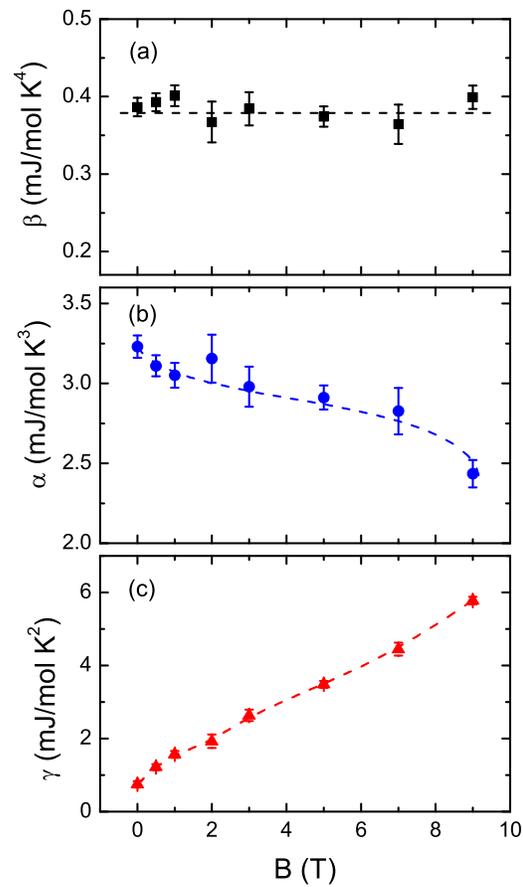}
\caption {(color online) Field dependence of the fitting parameters
$\beta$ and $\alpha$ and $\gamma$.} \label{fig4}
\end{figure}

The clear data and reasonable fitting parameters give us the
confidence about the credibility of the present analysis process.
More importantly, very few fitting parameters involve in this
approach. However, we can't obtain the precise location of the line
nodes, since SH measurements integrate the information from
different directions of the Fermi surface. Most likely, the
magnitude and structure of the gaps on different Fermi surfaces are
rather different. We must emphasize that such a multi-band does not
destroy the validity of our analysis results because the electronic
SH from the Fermi pockets with line nodes will prevail that from
fully gapped pockets in the low temperature limit.

In summary, we studied the low-temperature specific heat on the
Ba(Fe$_{1-x}$Ni$_{x}$)$_{2}$As$_{2}$ single crystal with $x$ =
0.058. Before measurements, the single crystals were carefully
annealed to improve the sample quality. We found a clear evidence
for the presence of $T^2$ term from the raw SH data, which is
consistent with the theoretical prediction for the superconductors
with line nodes. Our result is very similar to the previous reports
on the Co-doped system. Future investigations on other systems of
the iron-pnictide superconductors are needed to check whether it's a
common feature for the presence of line nodes on the overdoped sides
of the electron doped 122 system.

\begin{acknowledgments}
This work is supported by the Natural Science Foundation of China
(No. 11204338) and the ``Strategic Priority Research Program (B)" of
the Chinese Academy of Sciences (No. XDB04040300 and XDB04030000).
Gang Mu expresses special thanks to Grants-in-Aid for Scientific
Research from the Japan Society for the Promotion of Science (JSPS)
(Grant No. P10026).
\end{acknowledgments}

\end{document}